\documentclass{aa} 
\pdfoutput=1
\usepackage{graphicx}
\usepackage{amsmath}
\usepackage{natbib}
\usepackage{amsbsy}
\usepackage{subfig}
\usepackage{multirow}
\usepackage{textcomp}
\usepackage{setspace}
\usepackage{pslatex}
\usepackage{microtype}

\bibpunct{(}{)}{;}{a}{}{,}
 
\begin{document}

   \title{\textbf{Standing} sausage waves in photospheric magnetic waveguides}

   \author{
   		  I. Dorotovi\v{c},\inst{1}
   		  R. Erd\'{e}lyi,\inst{2}
   		  N. Freij,\inst{2}
   		  V. Karlovsk\'{y},\inst{3}
          \and
          I. M\'{a}rquez,\inst{4}
          }
   \institute{Slovak Central Observatory, P. O. Box 42,
              SK-94701 Hurbanovo, Slovak Republic\\
              \email{ivan.dorotovic@suh.sk}
         \and
			 Solar Physics \& Space Plasma Research Centre (SP$^{2}$RC), School of Mathematics and Statistics, University of Sheffield, Hicks Building, Hounsfield Road, Sheffield, S3 7RH, United Kingdom \\
			 \email{robertus@sheffield.ac.uk;n.freij@sheffield.ac.uk}
         \and
             Hlohovec Observatory and Planetarium, Sl\'{a}dkovi\v{c}ova 41,
             SK-92001 Hlohovec, Slovak Republic
             \email{astrokar@hl.cora.sk}    
         \and
             Instituto de Astrofísica de Canarias, E38205, La Laguna, Tenerife, Spain
             \email{imarquez@ull.es}
             }

   \date{Received ; accepted }

	\abstract{}
	{By focusing on the oscillations of the cross-sectional area and the total intensity of magnetic waveguides located in the lower solar atmosphere, we aim to detect and identify magnetohydrodynamic (MHD) sausage waves.}
	{Capturing \textbf{several high-resolution time series of mangetic waveguides} and employing \textbf{a} wavelet analysis\textbf{,} in conjunction with empirical mode decomposition (EMD)\textbf{,} makes the MHD wave analysis possible. For this paper, two sunspots and one pore (with a light bridge) were chosen as representative examples of MHD waveguides in the lower solar atmosphere.}
	{The \textbf{waveguides} display a range of periods from 4 to 65 minutes. \textbf{These} structures display in-phase behaviour between the area and intensity, presenting mounting evidence for sausage modes within these waveguides. The detected periods point towards standing oscillations.}
	{The presence of fast and slow MHD sausage waves has been detected in three different magnetic waveguides in the solar photosphere. Furthermore, these oscillations are potentially standing harmonics supported in the waveguides which are sandwiched vertically between the temperature minimum in the lower solar atmosphere and the transition region. \textbf{The relevance of} standing harmonic oscillations is that their exploitation by means of solar magneto-seismology may allow insight into the sub-resolution structure of photospheric MHD waveguides.}

	
	\titlerunning{On standing sausage waves in photospheric magnetic waveguides}
	\authorrunning{I. Dorotovi\v{c} et al.}
	\maketitle


\section{Introduction}

	Over the past decades, many oscillatory phenomena have been observed within a wide range of magnetic waveguides in the solar atmosphere \citep{banerjee,wang2011standing,2012ApJ745L18A,2012LRSP92A}. Sunspots and pores are just two of these many structures and they are known to display solar \textit{global} oscillations. See a recent review by e.g. \cite{pinter2011effects}.

	The commonly studied oscillatory periods in sunspots are 3 and 5 minutes. These oscillations are seen in intensity, line of sight (LOS) velocity and LOS magnetic field. The source of the 5-minute oscillation is a result of forcing by the 5-minute (\textit{p}-mode) global solar oscillation \citep{marsh2008p}, which forms the basis of helioseismology \citep{thompson2006magnetohelioseismology,pinter2011effects}. The 5-minute oscillations are typically seen in simple molecular and non-ionized metal lines, which form low in the umbral photosphere and are moderately suppressed not only in the penumbra, but also in the chromospheric atmosphere above the umbra \citep{OASO}. The cause of the 3-minute oscillations is still unknown but there are two main streams of theories: they could either be standing acoustic waves which are linked to the resonant modes of the sunspot cavity or they could be low-$\beta$ slow magneto-acoustic-gravity waves guided along the ambient magnetic field \citep{OASO}. The 3-minute oscillations are seen in plasma elements that form higher up, in the low chromosphere, and these are also moderately suppressed in the penumbra \citep{ORWS}.

	Magnetohydrodynamic (MHD) theory, when applied to a cylindrical magnetic flux tube, reveals that a variety of waves can be supported, four of which are often reported in various structures in the solar atmosphere. \textbf{Slow sausage (longitudinal)} \citep{2009SSRv..149...65D,wang2011standing}, fast kink \citep{2009A&A...497..265A,2009SSRv..149....3A}, fast sausage \citep{mcateer2003observational} and Alfv\'en (torsional) waves \citep{jess}, each of which affects the flux tube in a specific way. The sausage modes are of interest here; the sausage mode is a compressible, symmetric perturbation around the axis of a flux tube which causes density \textbf{perturbations} that can be identified in intensity images \citep{fujimura}. Furthermore, due to the fact that the wave will either compress or expand the flux tube, the magnetic field will also show signs of oscillations. This mode may come in two forms in terms of phase speed classification: a slow mode (often also called the longitudinal mode) which generally has a phase speed close to the characteristic tube speed, and, a fast mode, which has a phase speed close to the external sound speed, assuming a region that has a plasma-$\beta > 1$ \citep{2003ASSL..294.....G,rbook}. A main difference between the two modes is the phase relationship between appropriate MHD quantities which allows them to be identified. In this case, the fast sausage mode has an out-of-phase relationship between the area and intensity, while the slow sausage mode has an in-phase relationship. The technique that was applied to obtain these phase relationships are covered by  e.g. \cite{goedbloed,fujimura,moreels2013phase,michal2013}.

	Sausage modes have been observed in solar pores; \citet{doretala} observed a pore for 11 hours and reported periodicities in the range of 20-70 minutes. These oscillations were consequently interpreted as linear low-frequency slow sausage waves. \citet{morton} used the Rapid Oscillations in the Solar Atmosphere (ROSA) instrument to also identify linear sausage oscillations in a magnetic pore. However, determining whether the oscillations were slow or fast proved to be difficult. \citet{2012NatCo...3E1315M}, found the presence of fast sausage and kink waves higher in the solar atmosphere with sufficient wave energy to heat the chromosphere and corona.

	The source and driving mechanism(s) of these MHD sausage modes have been very difficult to identify. Numerical simulations of a flux tube rooted in the photosphere which is buffeted by a wide range of coherent sub-photospheric drivers is one method to identify \textbf{the} potential source of \textbf{MHD sausage waves}. These drivers can either be horizontal or vertical, single, paired or a power spectrum, with varying phase differences \citep[see e.g.][]{malins,khomenko,fedun2,fedun1,vigeesh2012three}. One example of a horizontal driver \textbf{is the absorption of the global solar \textit{p}-mode oscillation} by \textbf{a} sunspot \citep{1992ApJ...384..348G}. More recently, \citet{mathew} also studied this absorption and found a structured ring-like absorption pattern in Doppler power close to the umbral-penumbral boundary. This effect was largest where the transverse magnetic field was at its greatest and this region allows fast waves to be converted into slow magneto-acoustic waves, which are a potential source of MHD waves in sunspots and other similar magnetic structures.

	We report here, the observation of both \textit{slow} and  \textit{potentially} \textit{fast} sausage MHD waves in the lower solar atmosphere in three magnetic waveguides. In section \ref{sect2}, we describe the data collection and the data processing method. In section \ref{sect3}, we describe the results obtained from the three different data series and discuss \textbf{our} findings. Section \ref{sect4} details the underlying idea of identifying these oscillations as \textit{standing} harmonics. Finally, in section \ref{sect5}, we conclude.

\section{Data collection and Method of Analysis}
\label{sect2}

	Three time series of images with high angular resolution have been chosen here in order to demonstrate the identification of MHD sausage waves. The images were taken in the G-band ($430.5$ nm), which samples the low photosphere. This line forms deep in the photosphere and has a line intensity defined as $\rho{^2} \times$ line-of-sight column depth.
	
	The images were acquired using:

	\begin{figure}
		\centering
		\includegraphics[width=9cm,height= 6cm]{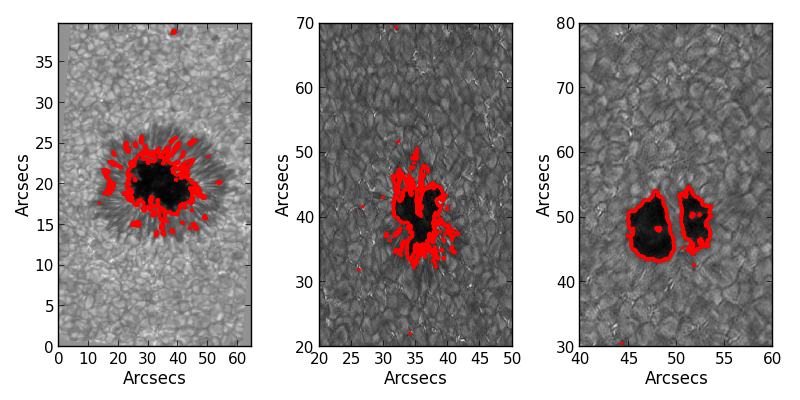}
		\caption{
		An overview of the magnetic waveguides observed for this analysis. \textit{(left)} The 1999 sunspot observed with the SVST with an average umbral area of 19,650 pixels (50 Mkm$^{2}$) . \textit{(middle)} The 2005 sunspot observed with the DOT with an average umbral area of 12,943 pixels (32 Mkm$^{2}$). \textit{(right)} The 2008 pore observed with the DOT with an average area of 10971 pixels (27 Mkm $^{2}$), the light bridge that separates the pore can be seen. Furthermore, these structures were seen near the disk centre, so there is little to no LOS effects. \textbf{The red line shows the thresholding technique applied to each waveguide at the start of the data series.}}
		\label{images}
	\end{figure}

	\begin{enumerate}
		\item 
			The Swedish Vacuum Solar Telescope (SVST) situated \textbf{on} La Palma \textbf{in} the Canary Islands. \citet{scharmer} provides a detailed description of the features of the SVST. The images were taken on the 7th July 1999. The sunspot is in the Active Region \textbf{(AR)} NOAA 8620. The observing duration is 133 minutes with a cadence time of 25 seconds. The field of view covers an area of $33,600$ km by $54,600$ km ($1$ pixel $\approx$ $60$ km). \citet{bonet} gives a detailed analysis of this sunspot. A context image is the left image of Fig. \ref{images}.
		\item 
			The Dutch Open Telescope (DOT) is also situated \textbf{on} La Palma \textbf{in} the Canary Islands. Two series of imaging data sequences were taken using this telescope. A detailed guide of the features of the DOT is  provided by \citet{rutten}. The first series of data were taken on the 13th July 2005; the sunspot is in the \textbf{AR} NOAA 10789. The region \textbf{slowly decayed} and this sunspot \textbf{was} leading \textbf{a} small group \textbf{of other magnetic structures}. The observing length is 165 minutes and has a cadence time of $30$ seconds. The second set of data, taken on the 15th October 2008, is of one large pore with a light bridge which is about 15 pixels (750 km) wide in the \textbf{AR} NOAA 11005. The duration of the observing run is 66 minutes and has a cadence time of $20$ seconds. Both DOT image sequences cover an area of $50,000$ km by $45,000$ km, where the maximum spatial resolution is 0.2" ($\approx$ $140$ km). Typical context images are the middle and right panels of Fig. \ref{images}.
   \end{enumerate}
   
	In order to obtain information relating to the cross-sectional area of these waveguides, a strict and consistent definition of the area is required. \textbf{This definition is that each pixel which has a value of less than $3\sigma$ of the median background intensity is counted as part of the waveguide}. The background is defined as an area of the image where there are no formed magnetic structures. This may appear to be an arbitrary definition, however, a histogram of the background intensity reveals a Gaussian distribution and when adding the area around and including the waveguide, there is significant peak on the lower end of the Gaussian distribution curve around 3$\sigma$ or higher. Thus, we have a $99\%$ confidence that the area is of the structure and not of the background.
	
	Fig. \ref{images} shows each waveguide at the start of the time series, where the red contour line represents the area found. The definition is accurate, however, it does include some non-waveguide pixels. The total intensity was determined by summing over the intensity of each pixel found in the waveguide. These waveguides are not static structures, they slowly changed in size during the observing period. This \textbf{background trend has} to be removed in order for it not to mask any weak oscillation signatures. The detrending was accomplished by a non-linear regression fit and the consistency of the results was compared to subtracting the residue from an Empirical Mode Decomposition (EMD) analysis (explained below). The residue is the data that remains after the EMD procedure has extracted as many signals as possible and it provides a very good approximation of the background trend.
 
	The resulting reduced data series were then analysed with a wavelet tool in order to extract any periods of oscillation  present within the data. The algorithm used is an adapted version of the IDL wavelet routine developed by \citet{torrence}. The standard Morlet-wavelet, which is a plane sine wave with an amplitude modulated by a Gaussian function, was chosen due to its suitable frequency resolution. The \textbf{white} cross-hatched area marks the cone of influence (COI), where edge effects due to the wavelet structure affect the wavelet transform and anything inside the COI is discarded. The \textbf{white dashed line contour} show the confidence level of 95\%. The wavelet method is very susceptible to noise at \textbf{short} periods and at times may not identify the true power of \textbf{short} periods.
	
	Further to this, the data representing the size and intensity has also been analysed using EMD, which decomposes the time series into a finite number of Intrinsic Mode Functions (IMFs). IMFs are essentially narrowband-filtered time series, with each IMF containing one or two periods that exist in the original data series. The EMD technique was first proposed by \citet{huang} and offers certain benefits over more traditional methods of analysis, such as wavelets or Fourier transforms. However, one drawback is that it is very prone to error with regards to \textbf{long} periods. For more information on the features and applicability of the EMD method see e.g. \citet{terradas}. The problems associated with both the wavelet and EMD process means that the two complement each other. Further, periods that appear in the wavelet \textit{just} below the confidence level, but appear strongly in the EMD process, is a good indication that a period is not spurious. Generally, the next step after EMD analysis is to construct a Hilbert power spectrum which has a better time and spatial resolution than either wavelet or FFT routines. However, this has not been carried out due to a lack of a robust code base at this time and will be addressed in future work. At this stage, we rely on wavelet and EMD analyses, as customary in solar physics.

\section{Results and Discussion}
\label{sect3}

\subsection{LOS, Circularity and Evolution of the Waveguide}

   \begin{figure*}
		\centering
		\subfloat{\includegraphics[height=6cm,width=18cm]{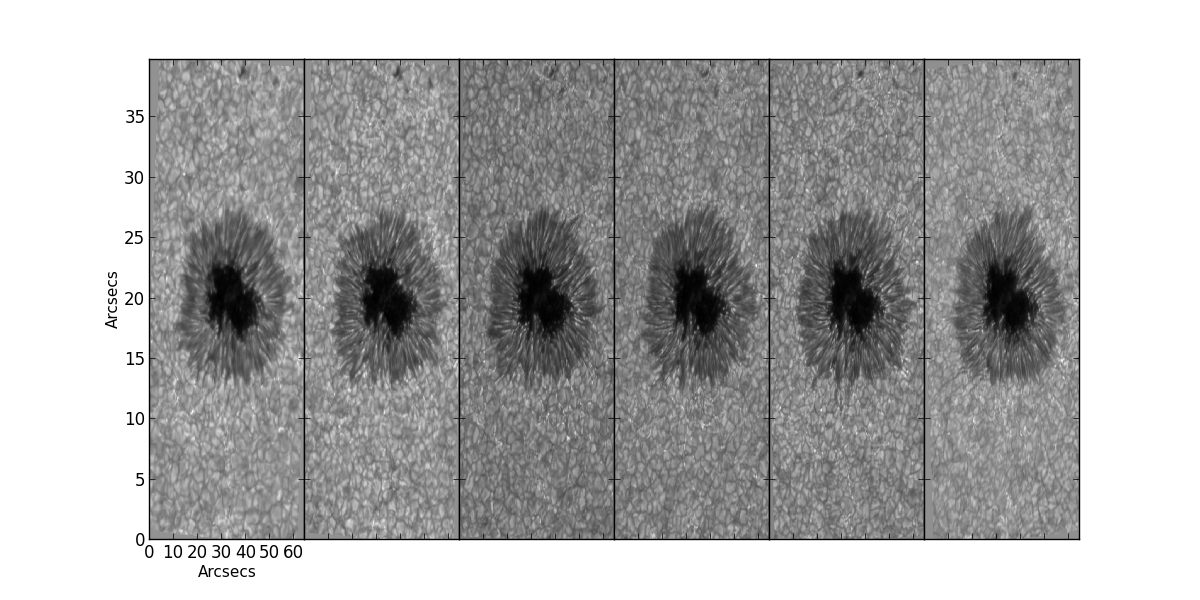}}\vspace*{-0.45cm}
	  	\subfloat{\includegraphics[height=6cm,width=18cm]{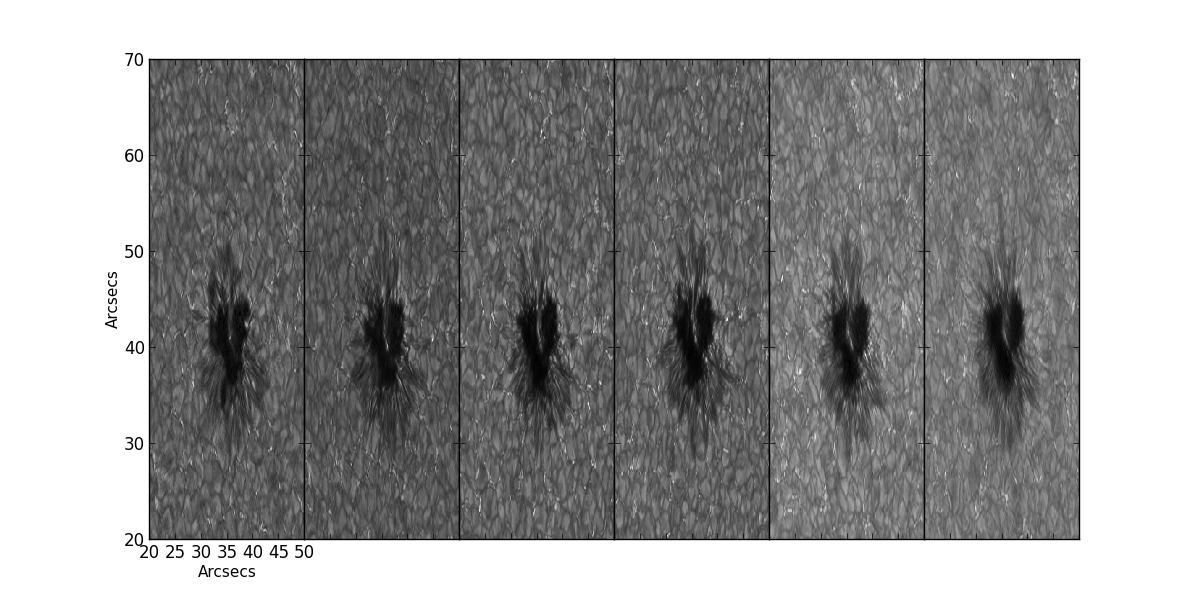}}\vspace*{-0.45cm}
	  	\subfloat{\includegraphics[height=6cm,width=18cm]{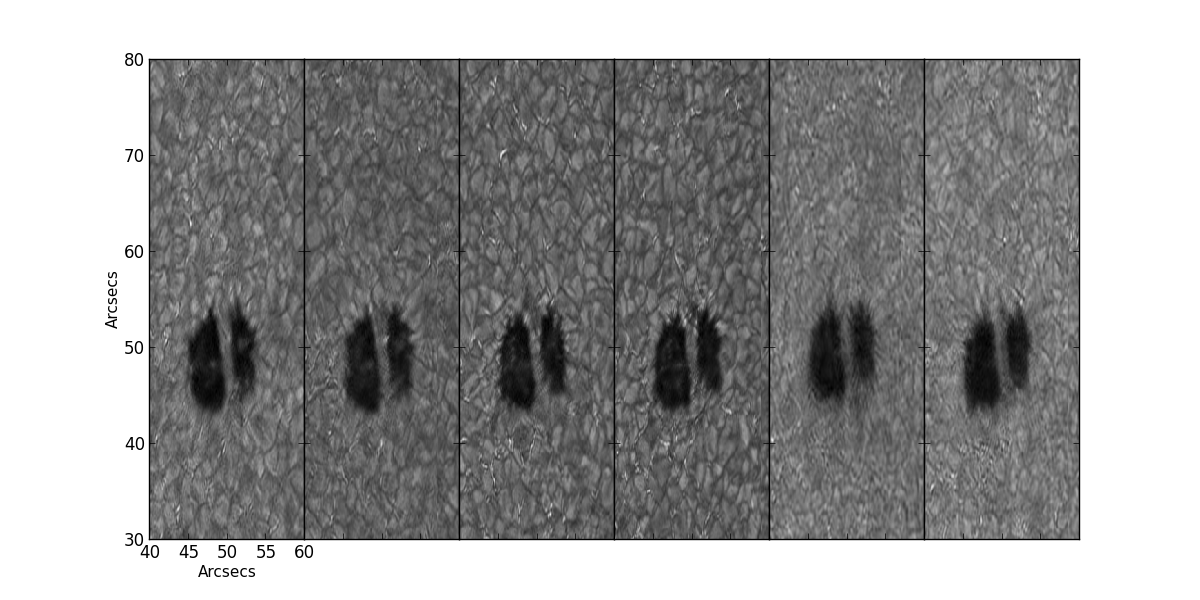}}
  		\caption{
	 	 		The waveguides are shown through six different parts of the observation sequence. The image sequence has time increasing from left to right. The first row is the 1999 sunspot, the middle row is the 2005 sunspot and the last row is the 2008 pore.
          		}
  		\label{data}
  \end{figure*}  
	
	Several points need to be clarified with regards to the data presented here before the full analysis.
	Firstly, LOS issues: \citet{2003A&A...397..765C,2003A&A...409..325C} investigated how the LOS angle affects various aspects of observing coronal loops in a 2D model.
	Overall they found that for the slow sausage MHD wave, for a range of angles from $\pi/6$ to $\pi/3$, the observed intensity decreases as the LOS angle increases.
	Further, the larger angles lengthened the \textit{observed period} of the wave.
	While the objects here are not coronal loops, the LOS angle still matters and should behave similarly.
	The LOS angles in all three cases were less than 30$^\circ$ limiting any relevant effects due to LOS.
	
	Sunspots or pores are not fully circular and can have arbitrary shapes.
	The effects of non-circular shape have been studied by, for example, \citet{2003A&A...409..287R,2009A&A...502..315M,2011A&A...527A..53M}.
	While they do not account for the very complicated and real structure of the sunspots and pores observed here, they still offer an adequate insight.
	Current theory suggests the shape will have a minor effect on the oscillations unless it has a significant deviation from circularity.
	Further, the structure of each waveguide undergoes minor change during the observation campaign, limiting any effects from large-scale structural change, as can be seen in Fig. \ref{data}.

\subsection{MHD Theory for Phase Relations}
	
	Treatment of the MHD equations makes it possible to determine phase relations between various physical quantities for propagating and standing MHD waves. This has been summarised briefly by \citet{goedbloed} and also applied by \citet{fujimura}. \citet{fujimura} found that the phase relation for the slow MHD wave with regards to cross-sectional area and density to be in-phase regardless \textbf{of whether} the wave is propagating or standing. More recently, \citet{moreels2013phase} expanded on this idea, taking into account factors such as LOS which were neglected earlier, but also expanding the theory to cover fast MHD sausage waves. The phase relation for the magnetic field to the cross-sectional area is in-phase assuming that the plasma is frozen-in to the magnetic field.
	
	 Supplementary information from other perturbation phase relations, such as velocity and the magnetic field, allows one to determine whether the observed MHD wave is slow or fast. In summary, the slow MHD sausage mode has an in-phase behaviour between intensity and area perturbations, while the fast sausage mode has an out-of-phase behaviour. Before progressing, we need to address the opacity effect on MHD wave perturbations. This is relevant, since intensity fluctuations can be due to the change of the optical depth along the LOS, which has the same phase \textbf{difference} as the fast MHD sausage wave and as a result is indistinguishable without further information \citep{fujimura}.
	
	Recently, \citet{michal2013} determined analytically the phase difference between the cross-sectional area and the total intensity perturbations for both the slow and fast MHD sausage modes.
	They found that, for both the slow body and surface MHD wave, the behaviour is in-phase, while for the fast surface wave, the behaviour is out-of-phase.
	This result means that it is possible to approximately separate slow and fast sausage waves without the use of other observable variables. Their results will be used here in-order to distinguish between slow and fast MHD sausage modes.

\subsection{Sunspot, 7 July 1999 , AR 8620}

   \begin{figure*}
	   \centering
	   \subfloat{\includegraphics[width=9cm]{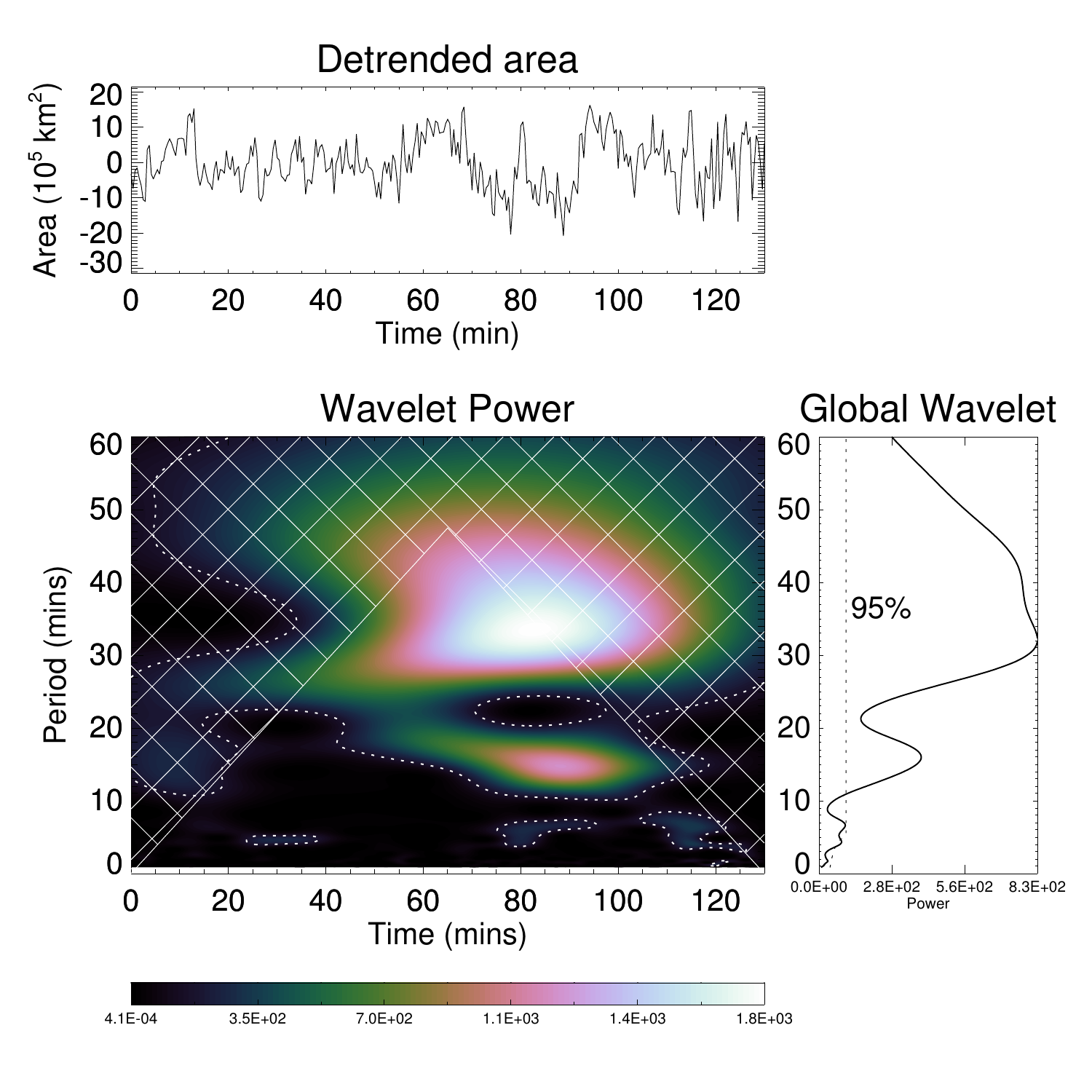}}
	   \subfloat{\includegraphics[width=9cm]{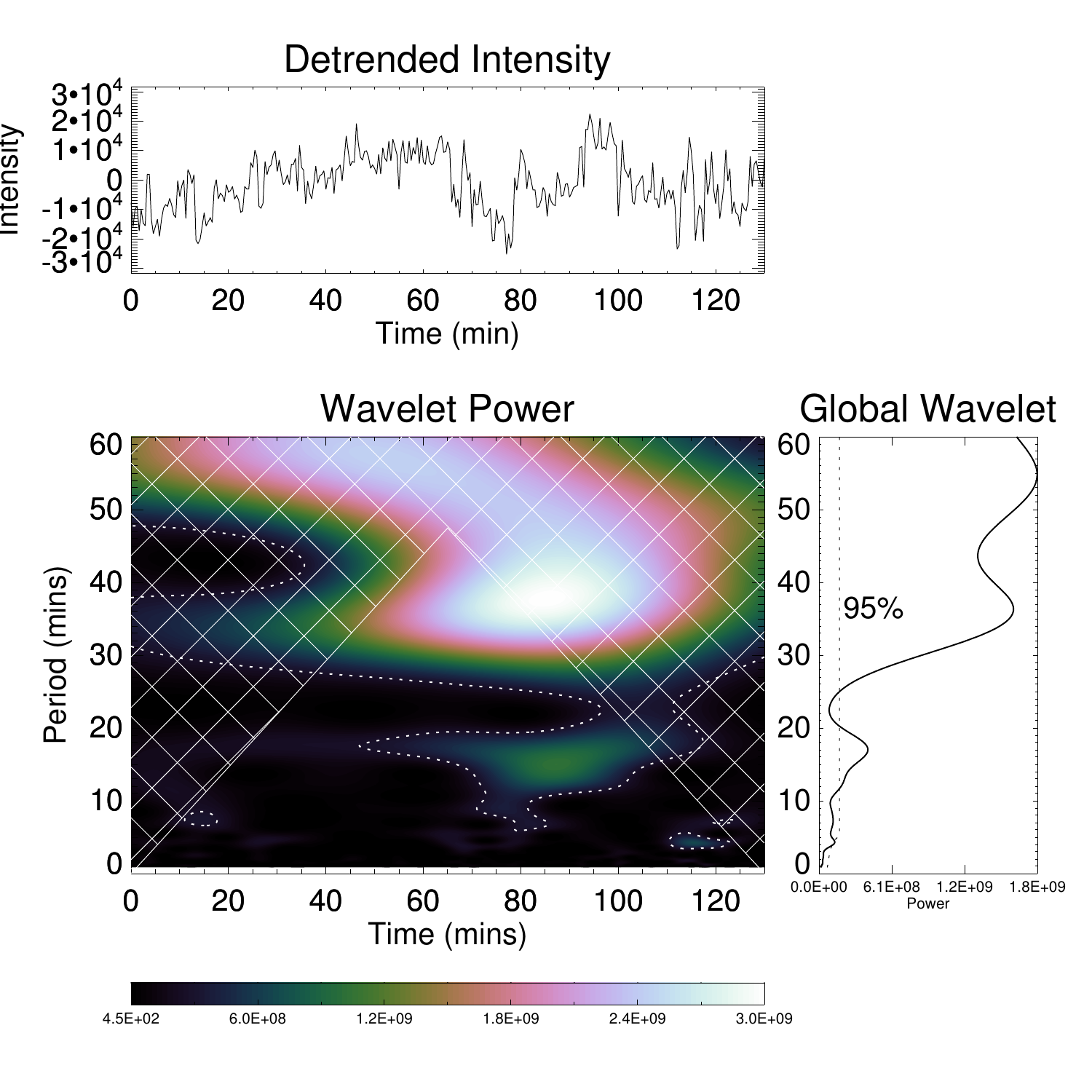}}
	   \caption{
				\textit{(left image)} Evolution of the area of the 1999 sunspot (\it{upper panel}); \textnormal{the wavelet power spectrum for a white noise background, the cone of influence is marked as a cross-hatched area where edge effects become imporant and the contour lines show the 95\% confidence level} (\it{lower left panel}). \textnormal{Global (integrated in time) wavelet power spectrum, where the dashed line shows the 95\% confidence limit} (\it{lower right panel}).  \textit{(right image)} \textnormal{The same as the left image but for the mean intensity of the 1999 sunspot.} 
				}
	   \label{1999sunspot}
   \end{figure*}
   	
	Fig. \ref{1999sunspot} shows the wavelet analysis of the 1999 sunspot area and intensity data. There are four confidently identified periods that exist in the area wavelet with 95\% certainty; 4, 7, 16 and 32 minutes. The 32-minute period is found over a wide range of the time series, with some of its power inside the COI. However, most is confidently outside the COI. The 16-minute period is strongly localised at 50 to 120 minutes of the data series \textbf{and starts at 18 minutes and slowly} increases and stabilizes at 14 minutes. There is a third and fourth period at 4- and 7-minutes that just reaches the significance level and appear sporadically during the time series.
	
	The intensity wavelet shows three distinct periods of oscillations above the confidence level: 4, 16 and 36.5 minutes. The 36.5-minute period has a corresponding area wavelet oscillation at 32-minutes. While the 16-minute oscillation corresponds to the 16-minute oscillation found in the area. Further, the 16-minute period starts with its power very concentrated and does not display the same period change as the area oscillation does. Finally, the 4-minute period also corresponds to an oscillation found in the area but is also sporadic in its appearance.
	
	It is safe to say that these oscillations are caused by sausage waves. The reason is that in linear ideal MHD theory, the sausage wave is the only MHD wave capable of changing the area of the flux tube that is observed on disk \citep[see e.g.][]{2003A&A...397..765C,wang2004coronal}. Without the ability to directly compare the phase difference of the area to \textbf{the} intensity, great caution needs to be exercised to determine with confidence whether the perturbations are fast or slow. A wavelet phase diagram reveals regions (where the wavelet coherence is high and the period is $\le 20$ minutes) to be either out-of-phase or in-phase but a clear image of constant phase \textbf{difference} does not appear. This might be due to mode conversion occurring in the sunspot, since the G-band samples a region where the plasma-$\beta$ $\approx 1$ in a magnetic structure \citep{gary}. When the period is $\ge 20$ minutes, the only area of high coherence is located around 30 minutes and found to \textbf{be nearly} out-of-phase, which hints that there might be a fast \textbf{surface} sausage wave. However, only two full wave periods are outside the COI, which is due to the total length of the data series. This behaviour indicates that for short periods, a mixture of fast \textbf{surface} and slow MHD sausage waves are present while for the \textbf{long} period, it is purely a fast \textbf{surface} MHD sausage wave.

	\begin{figure*}
	\centering
	\includegraphics[width=18cm,height=11cm]{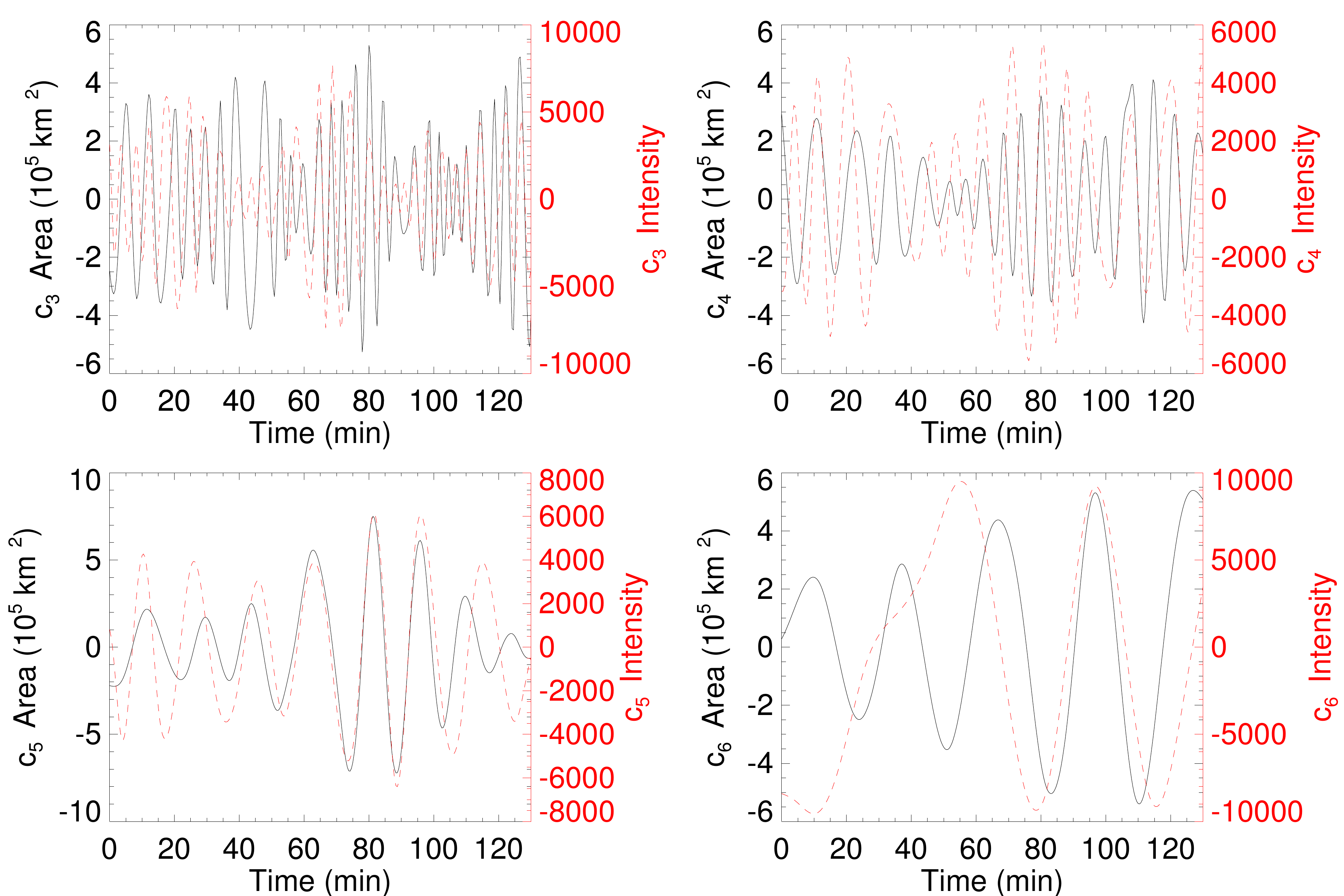}
    \caption{
	      			The IMFs of the evolution of the area (red) and intensity (black) for the 1999 sunspot, over-plotted to aid comparison. Generally after the $6$th IMF, higher IMFs lack a sufficient number of wave periods, which makes it difficult, and less reliable, to obtain an accurate period.
   		  }
    \label{1999IMF}
	\end{figure*}
		
	Fig. \ref{1999IMF} shows the computed IMFs for the 1999 sunspot data set. The IMFs show the periods of oscillations identified using the EMD routine. Several IMFs are generated by the algorithm and IMFs which show irrelevant periods or the additional residue are ignored. In general, the higher order IMFs tend to show longer periods and as such contain fewer wave periods, which makes phase identification less reliable. Four IMF overlays are shown and IMFs with similar periods as the wavelet plots have been overlaid in order to aid comparison for each dataset.
	
	Four IMFs directly coincide with the wavelet period that reveal both area and intensity perturbations. IMF $c_{3}$ displays the 4-minute period where major regions of in-phase behaviour can be seen, however, either side shows one or two wave periods of out-of-phase behaviour. IMF $c_{4}$ \textbf{exhibits} a period of 7-minutes. The picture here is more muddled as an extra period is present in the intensity, namely 11 minutes, making phase identification harder for the 7-minute period. Where the IMFs coincide with the same period, namely at the start of the time series, the phase \textbf{difference} is approximately 45 degrees\textbf{, which} the authors have no theoretical explanation for. IMF $c_{5}$ displays a 16-minute period, with an in-phase behaviour. Finally, IMF $c_{6}$ contains the 32-minute period. This period does not fully match the period seen in the intensity but also one of the edge effects of the EMD process can be seen in the intensity signal. Near the end of the time series, the two IMFs overlap with the same period with an in-phase behaviour. In summary, the EMD process shows that the major behaviour is in-phase indicating the existence of a slow sausage mode. Also \textbf{the} regions of changing phase \textbf{difference} at lower periods indicates the potential existence of a fast surface mode. However, the last IMF does not agree with the wavelet phase due to the artefact from the EMD process.
	
	It was possible to approximately separate the penumbra from the umbra and investigate its area for oscillations. However, the penumbra is a highly dynamic object and this makes the area estimation reasonably uncertain. There seem to be four periods that exist at 95 \% certainty: 5, 9, 15 and 25. The three shorter periods (5, 9 and 15 minutes) closely correspond to the 4-, 7- and 16-minute oscillations in the umbra; they could be a continuation of these umbral periods that became up-shifted as they enter the less compact structure of the penumbra. While the 25-minute period does not directly correspond to an observed area oscillation. The wavelet phase analysis shows large regions of out-of-phase behaviour where the period is either below 10-minutes or above 20-minutes. This behaviour is a mixed collection of fast \textbf{surface} and slow sausage modes, with regions moving from one phase \textbf{difference} to another after 3 or more wave periods.

\subsection{Sunspot, 13 July 2005, AR 10789}
   
   \begin{figure*}
   \centering
   	   \subfloat{\includegraphics[width=9cm]{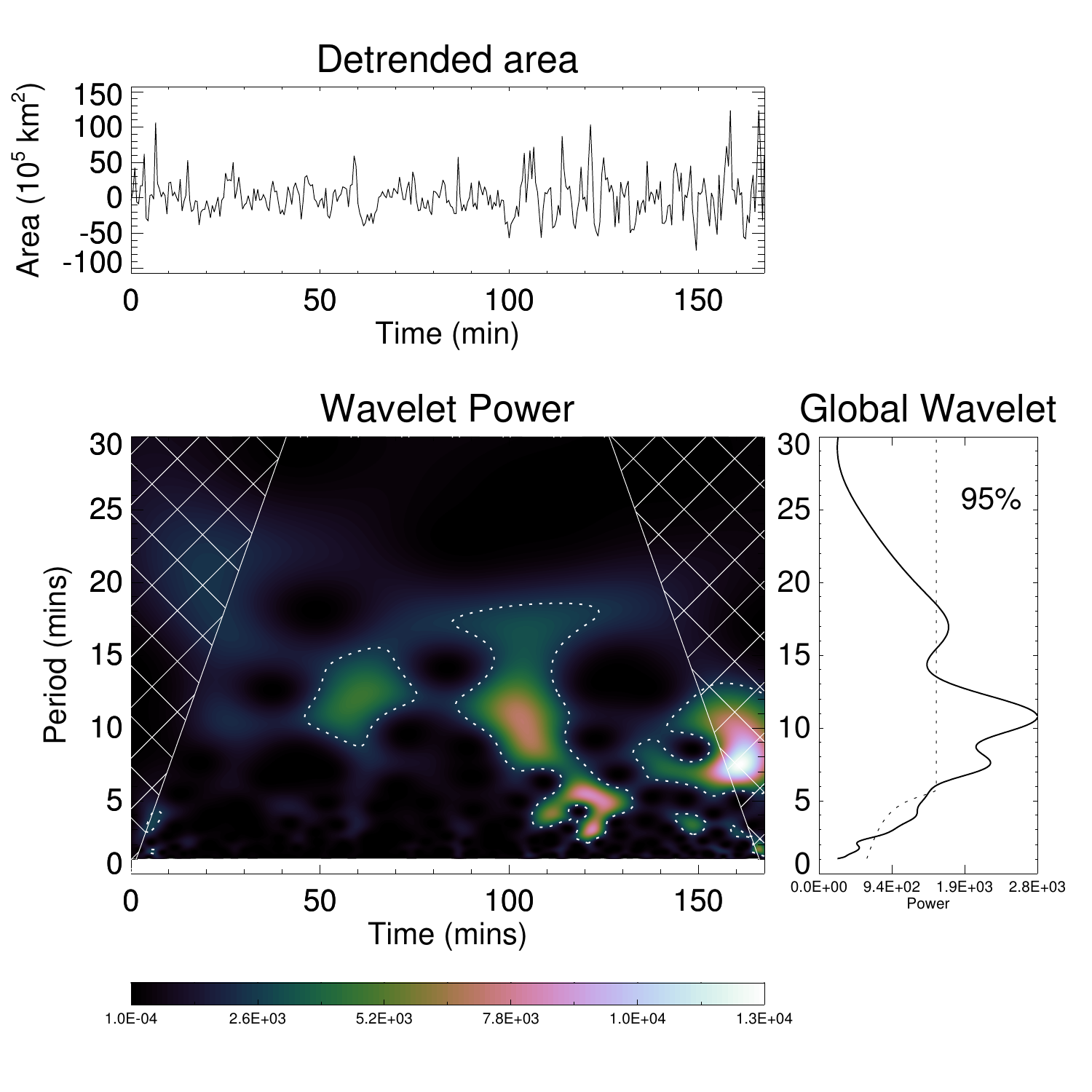}}
   	   \subfloat{\includegraphics[width=9cm]{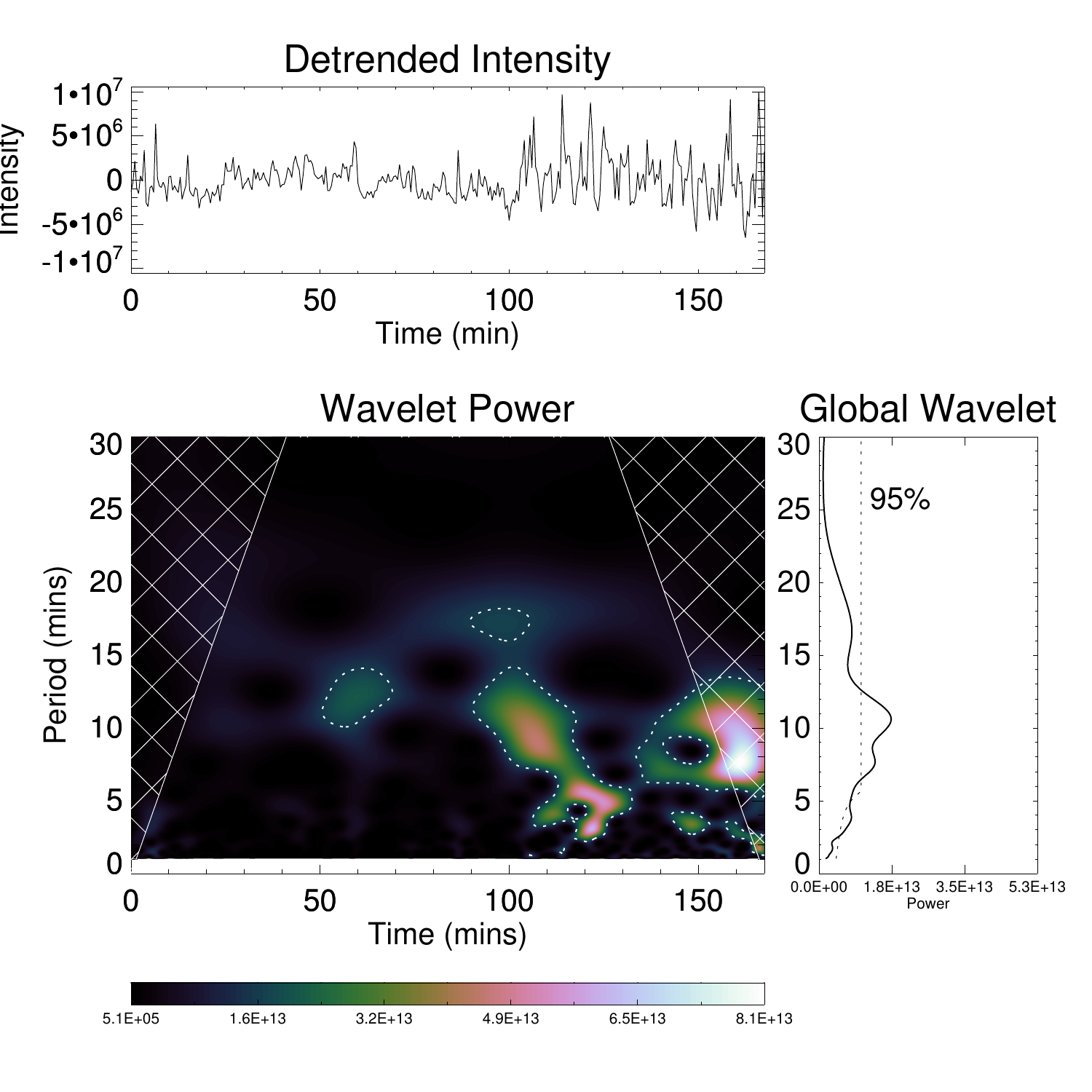}}
      \caption{
      			Same as Fig. \ref{1999sunspot} but for the sunspot in \textbf{AR} 10789 in 2005.
      		  }
      \label{2005sunspot}
   \end{figure*}

   \begin{figure*}
   \centering
   \includegraphics[width=18cm,height=11cm]{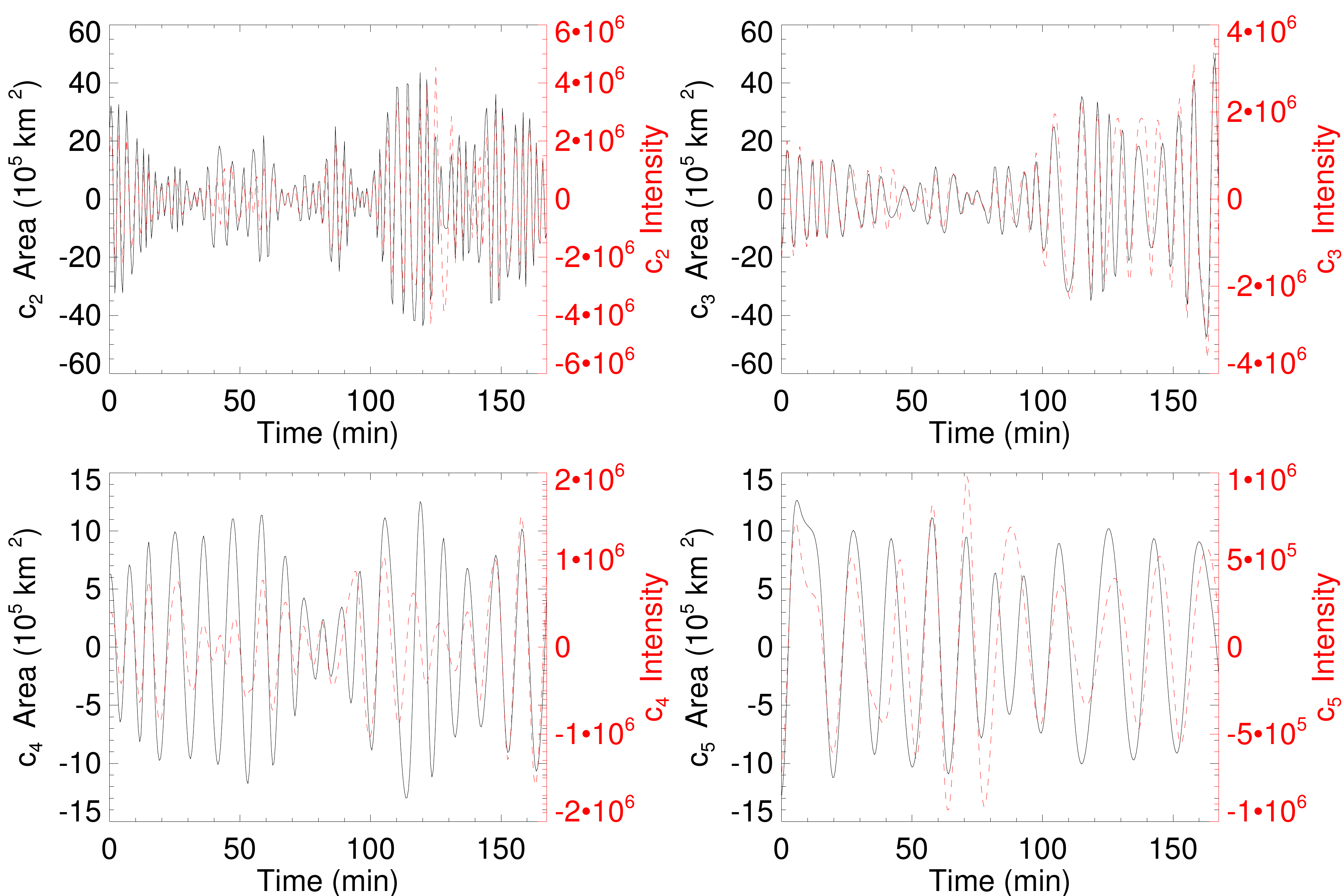}
      \caption{
      			Same as Fig. \ref{1999IMF} but for the sunspot in \textbf{AR} 10789 in 2005.
      		  }
      \label{20005IMF}
   \end{figure*}
   
	Fig. \ref{2005sunspot} shows the wavelet analysis of the 2005 sunspot area and intensity in AR 10789. There are four periods that exist at 95 \% confidence level: 4, 7.5, 11 and 16.5 minutes. Each period has a region of high power in the wavelet, with the lower periods appearing nearer the end of the time series. The corresponding intensity wavelet reveals that there are three periods: 4, 7.5 and 10.5 minute oscillations, however, the 16.5-minute oscillation is present but is a very weak signal. The cross-wavelet phase indicates that these oscillations are in-phase. There are no major regions of out-of-phase behaviour.
		
	Fig. \ref{20005IMF} shows the IMFs for the area and the intensity of the sunspot data in AR 10789. In this case, each period is found by the EMD process. IMF $c_{2}$, IMF $c_{3}$, IMF $c_{4}$ and IMF $c_{5}$ correspond to the 4, 7.5, 11 and 16.5-minute oscillation \textbf{periods} respectively. IMF $c_{2}$ displays extensive in-phase behaviour throughout the time series which is a strong indication of the slow sausage MHD wave at a period not too dissimilar to the global \textit{p}-mode oscillation. The region of interest is within the time interval of 90-130 minutes for IMF $c_{4}$, where the wavelet has these oscillations. The IMF shows clear in-phase behaviour in this time interval. The overall phase relation between the area and intensity indicates the presence of slow sausage waves.
	
\subsection{Pore, 15 October 2008}

   \begin{figure*}
   \centering
    \subfloat{\includegraphics[width=9cm]{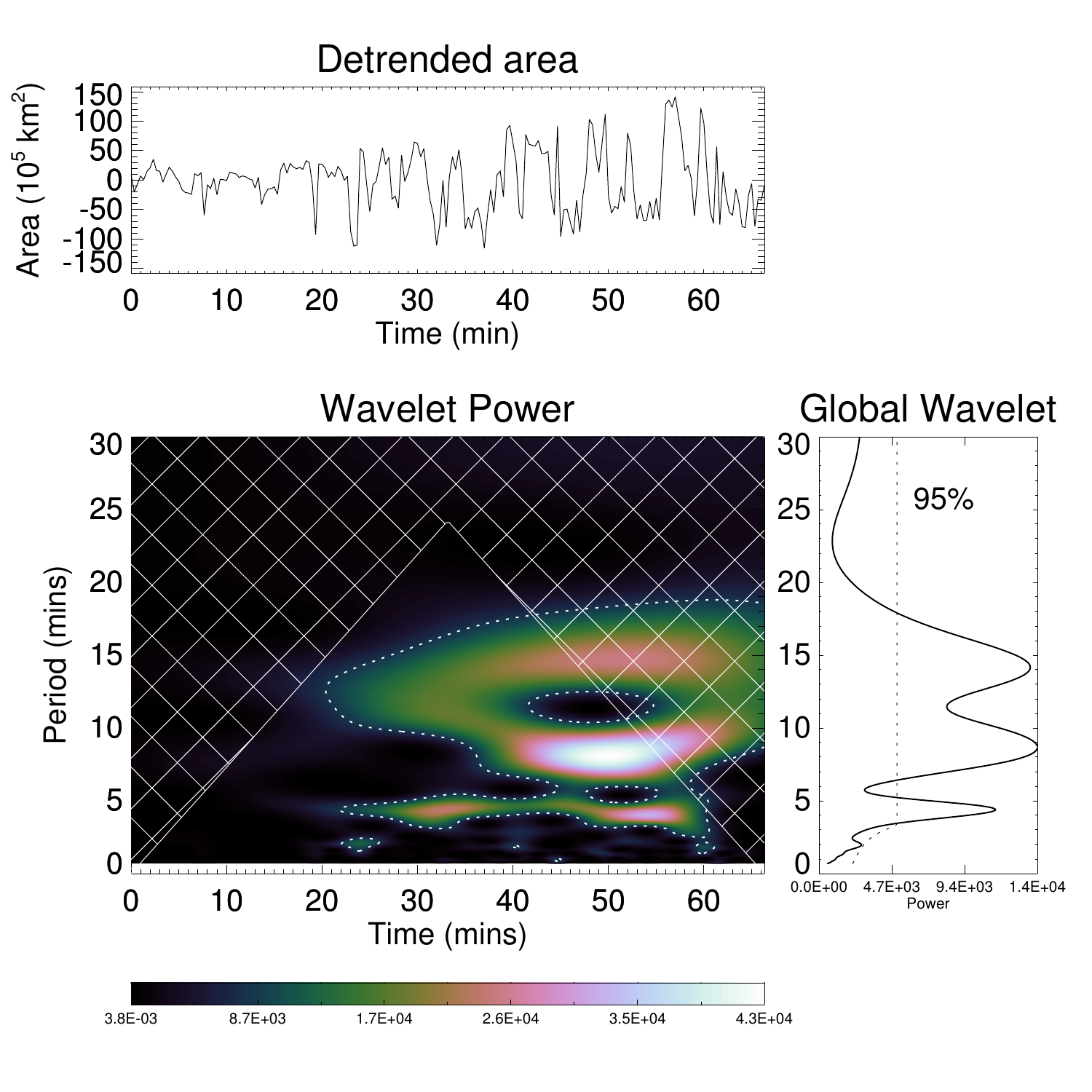}}
   	\subfloat{\includegraphics[width=9cm]{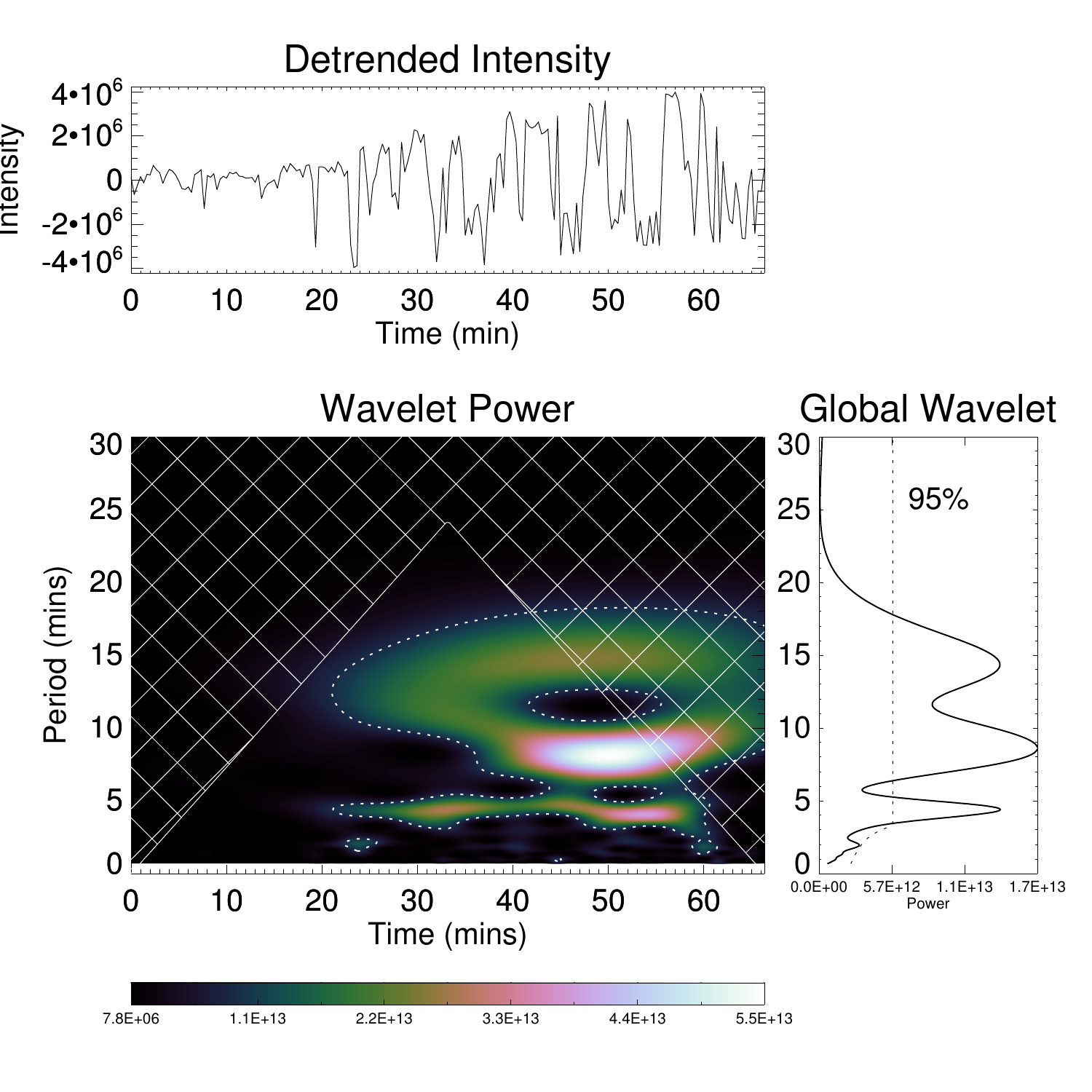}}
   	   \caption{
      			Same as Fig. \ref{1999sunspot} but for the pore in \textbf{AR} 11005 in 2008.
 		      }
      \label{2008pore}
   \end{figure*}
   
     \begin{figure*}
     \centering
     \includegraphics[width=18cm,height=11cm]{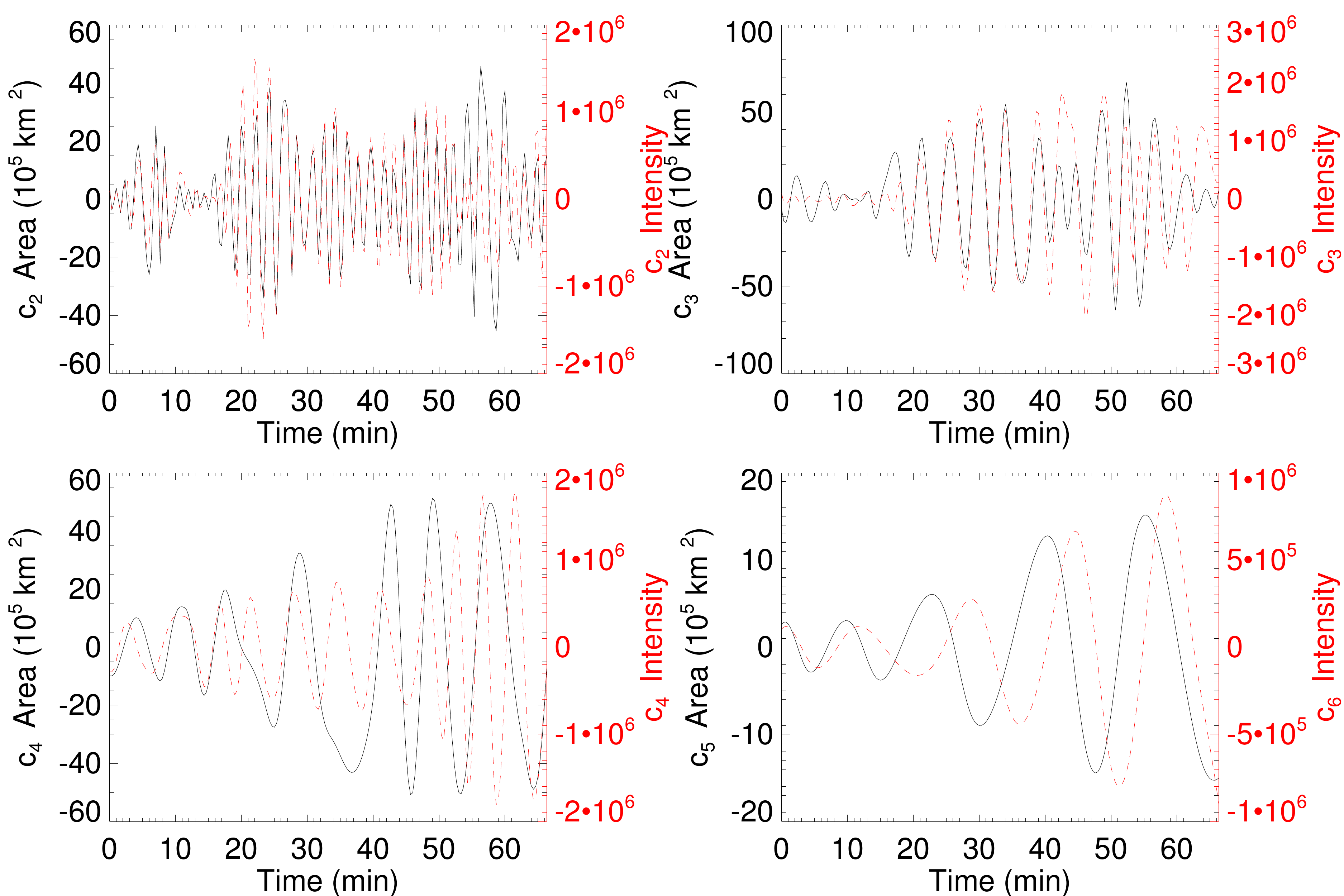}
     \caption{
       			Same as Fig. \ref{1999IMF} but for the pore in \textbf{AR} 11005 in 2008.
        		  }
     \label{20008IMF}
     \end{figure*}
      
	Fig. \ref{2008pore} shows the wavelet analysis of the pore with a light bridge. There are three periods that exist at 95 \% confidence level: 4.5, 8.5 and 14.5 minutes. The large part of the power of the period of 14-15 minutes is inside the COI, however, the period appears in the EMD analysis and has a large fraction of power outside the COI and thus has not been ignored for this analysis. The three periods are seen in both area and intensity data when the wavelet analyses are cross-correlated. The power for these two periods is concentrated in the time interval of 20-60 minutes. The cross-wavelet analysis shows that the overlapping time span is somewhat smaller, at about 30-50 minutes. Further, the wavelet power for each period runs parallel to each other throughout the time series, they appear at the same time \textbf{and} seem to fade away at a similar time as well.
	
	Fig. \ref{20008IMF} shows the IMFs for the area with intensity over-plotted. In this case, IMF $c_{3}$ indicates a period of 4.5 minutes and IMF $c_{4}$ has a characteristic period of 8.5 minutes; this applies to both the area and intensity IMFs. IMF $c_{3}$ reveals that the phase relation is in-phase for the majority of the time series. IMF $c_{4}$, also reveals large regions of roughly in-phase behaviour but with, again, a 45 degree phase difference. Not shown is the comparison of IMF $c_{4}$ and IMF $c_{5}$ for the area and intensity, respectively. At the end of the time series for both, there is a mixture of in-phase and intensity leading the area for the 8.5 minute oscillation. IMF $c_{5}$ and IMF $c_{6}$ for the area and intensity, respectively, show a period of 14.5 minutes. There is a region of near out-of-phase behaviour before this then turns into 45 degree phase \textbf{difference} with the area leading the intensity perturbations. Consistently, there are occurrences of unexplainable phase differences which require theory to be developed to explain.
       
	The easiest way to confirm the linearity of waves is to compare the amplitude of the oscillations to the characteristic scale of the structure. In all three cases studied here, the oscillation amplitudes are around 10\% or less of the total area, which indicates that these oscillations are linear. Furthermore, the amplitude of the oscillation in the last two cases is \textbf{by and large} the same, so the amplitude \textbf{has} scaled with the size of the structure. However, for the 1999 sunspot, the amplitude of the oscillation is an order of a magnitude less. Whether this is due to the large size of the sunspot or the very stable nature during the observation window needs to be investigated in future work.
	
\section{Standing Harmonics}
\label{sect4}
	\begin{table}
	\centering
	\begin{tabular}{ccc}
		\hline
		Data Set & Period (Mins) & Ratio ($P_{1}/P_{i}$) \\ \hline \hline
		\multirow{3}{*}{Sunspot 1999} & $P_{1}$ - 32 $\pm$ 2.5 & - \\
							  		  & $P_{2}$ - 16 $\pm$ 1.5 & 2 $\pm$ 0.2 \\
							  		  & $P_{3}$ - 7 $\pm$ 0.5 & 4.6 $\pm$ 0.3 \\
							  		  & $P_{4}$ - 4 $\pm$ 0.5 & 8 $\pm$ 0.5 \\ \hline
		\multirow{3}{*}{Sunspot 2005} & $P_{1}$ - 16.5 $\pm$ 1.5  & - \\
					      			  & $P_{2}$ - 11 $\pm$ 0.5 & 1.5 $\pm$ 0.2 \\
					      			  & $P_{3}$ - 7.5 $\pm$ 0.5 & 2.2 $\pm$ 0.2 \\
					      			  & $P_{4}$ - 4 $\pm$ 0.5 & 4.2 $\pm$ 0.6 \\ \hline
		\multirow{3}{*}{Pore 2008}    & $P_{1}$ - 14.5 $\pm$ 0.5 & - \\
		 							  & $P_{2}$ - 8.5 $\pm$ 0.5 & 1.7 $\pm$ 0.1 \\
					      			  & $P_{3}$ - 4.5 $\pm$ 0.5 & 3.2 $\pm$ 0.2 \\ \hline
	\end{tabular}
		\caption{The periods of oscillations that are found in the area of the waveguides that exist at 95\% confidence level.}
		\label{harm_table}
	\end{table}
		
	Basic MHD theory interpretation allows sunspots and pores to be described as vertical cylindrical flux tubes, with the base bounded in the photosphere and the top bounded at the transition region due to the sharp gradients in the plasma properties at these locations. Taking this further, an ideal flux tube is assumed here. The plasma density and magnetic field are homogeneous within the flux tube. This means that the standing harmonics of such flux tubes are the MHD equivalent to those of the harmonics in an open-ended compressible air pipe, where the ratio of the harmonic periods is given by, \, $P_{1}/P_{2}=2, \,\, P_{1}/P_{3}=3$ and so forth. \textbf{This only applies in the long wavelength or thin tube approximation.} Using harmonic ratios to carry out magneto-seismology has been used for example, by \citet{2005ApJ...624L..57A,2005A&A...430.1109A} who researched the effects of longitudinal density stratification on kink oscillations and resonantly damped kink oscillations, while \citet{luna-cardozo} studied longitudinal density effects and loop expansion on the slow sausage MHD wave. \citet{luna-cardozo}, found that specific density profiles in lower atmospheric flux tubes could increase or decrease the value \textbf{of} the period ratio. The authors are unaware of any work that details the changes to further harmonic ratios, so the assumption that the amount of deviation from the canonical value for the period ratio ($P_{1}/P_{2}$) is the same for other period ratios e.g $P_{1}/P_{3}$ or higher is used.
	
	Let us now summarise the observed findings. Table \ref{harm_table}. contains the periods of oscillations found in all three magnetic waveguides. 
	
	For the 1999 sunspot, there are four periods found. The second period at 16 minutes gives a period ratio ($P_{1}/P_{2}$) of 2 $\pm$ 0.2, which is exactly the same as the expected value of a uniform waveguide with a canonical value of 2. The next period ratio is at 4.6 $\pm$ 0.3. Here, the change from canonical value is substantial if this is indeed the third period which should be around 10.6 minutes, unless the effect on the harmonic ratio increases with each successive ratio. The last period is difficult to incorporate into the harmonic standpoint and it is most likely that the 4-minute period is due the global \textit{p}-mode.
	
	For the 2005 sunspot in AR 10789, there is a clearer picture of potential harmonics. The first period is at 16.5 minutes and the second period is at 11 minutes, which gives a ratio of 1.5 $\pm$ 0.2, and the third period at 7.5 minutes gives a ratio of 2.2 $\pm$ 0.3. The period ratio is modified downwards in a consistent manner as the harmonic number increases. These ratios are strong evidence for standing waves in this magnetic waveguide. As was the case for the 1999 Sunspot, the period at 4 minutes has a period ratio that does not fit into this harmonic viewpoint and is most likely due to the global \textit{p}-mode instead.
	
	For the 2008 pore of AR 11005, the picture is more muddled due to the short time series available. Taking the 15-minute period to be the first harmonic, the ratio is 1.7 $\pm$ 0.1 for the 8.5-minute period, very similar to both first period ratios of the previous sunspots. The third period is again very close to the period of the global \textit{p}-mode.
	
	The main conclusion to take away from this data analysis so far is that the simple homogeneous flux tube model cannot fully account for these ratios. However, this simple model seems to be robust enough to give a good first insight. The most likely reasons for deviation from the canonical period ratio value are firstly that sunspots and pores (just like most lower atmospheric magnetic structures) expand with height, causing magnetic stratification \citep{2008A&A...486.1015V,luna-cardozo}, and secondly, that the Sun's gravity causes density stratification \citep{2009SSRv..149....3A}. These two effects will either increase or decrease the period ratio of the harmonics depending on the chosen density or magnetic profile \textbf{(see \citet{luna-cardozo} for a detailed analysis in the context of slow sausage oscillations or see \citet{2013SoPh..tmp..195E} for kink modes)}. In addition, these magnetic structures are rarely purely cylindrical, they can be elliptical (or arbitrary) in shape \citep[see][]{ruderman2009transverse,2009A&A...502..315M} and in most cases are non-axially symmetric. Also, in some cases the flux tube is more suitably described as closed-ended at the photosphere while open-ended at the transition region, which would remove the even harmonics.
	 
\section{Conclusions}
\label{sect5}

	In this paper we have investigated three magnetic waveguides, with the objective of detecting MHD sausage waves and determining whether they are slow or fast, propagating or standing. Based on the results presented here, we have confidently interpreted the observed periodic changes in the area cross section of flux tubes, which are manifested as a pore and two sunspot waveguide structures, as proof of the existence of linear slow and fast \textbf{surface} sausage MHD oscillations. Using wavelet analysis, we found standing waves in the photosphere with periods ranging from 4 to 32 minutes. Employing complementary EMD analysis has allowed the MHD modes detected to be identified as a combination of \textit{fast surface sausage} and \textit{slow sausage} modes, due to the phase \textbf{difference} of the area and intensity. It is very likely that these oscillations are \textit{standing harmonics} supported in a flux tube. The period ratio ($P_{1}/P_{i=2,3}$) of these oscillations indicates strongly that they are part of a group of standing harmonics in a flux tube that is non-homogeneous and is bound by the photosphere and the transition region. Furthermore, there is possible indirect evidence of mode conversion occurring in one of these magnetic waveguides.
	
\begin{acknowledgements}
	The authors thank J. Terradas for providing the EMD routine used in the data analysis and acknowledge M. Moreels for his theoretical discussions on phase relations. The authors would also like to thank the unknown referee for the helpful and insightful comments and suggestions. RE acknowledges M. K\'eray for patient encouragement and is also grateful to NSF, Hungary (OTKA, Ref. No. K83133). This work is supported by the UK Science and Technology Facilities Council (STFC). Wavelet power spectra were calculated using a modified computing algorithms of wavelet transform original of which was developed and provided by C. Torrence and G. Compo, and is available at URL: http://paos.colorado.edu/research/wavelets/ The DOT is operated by Utrecht University (The Netherlands) at Observatorio del Roque de los Muchachos of the Instituto de Astrofsica de Canarias (Spain) funded by the Netherlands Organisation for Scientific Research NWO, The Netherlands Graduate School for Astronomy NOVA, and SOZOU. The DOT efforts are part of the European Solar Magnetism Network. The SVST was operated by the Institute for Solar Physics, Stockholm, at the Observatorio del Roque de los Muchachos of the Instituto de Astrofísica de Canarias (La Palma, Spain)
\end{acknowledgements}

\bibliographystyle{aa}
\bibliography{ssw2011aa}

\begin{thebibliography}{49}
\expandafter\ifx\csname natexlab\endcsname\relax\def\natexlab#1{#1}\fi

\bibitem[{{Andries} {et~al.}(2005{\natexlab{a}}){Andries}, {Arregui}, \&
  {Goossens}}]{2005ApJ...624L..57A}
{Andries}, J., {Arregui}, I., \& {Goossens}, M. 2005{\natexlab{a}}, \apjl, 624,
  L57

\bibitem[{{Andries} {et~al.}(2009{\natexlab{a}}){Andries}, {Arregui}, \&
  {Goossens}}]{2009A&A...497..265A}
{Andries}, J., {Arregui}, I., \& {Goossens}, M. 2009{\natexlab{a}}, \aap, 497,
  265

\bibitem[{{Andries} {et~al.}(2005{\natexlab{b}}){Andries}, {Goossens},
  {Hollweg}, {Arregui}, \& {Van Doorsselaere}}]{2005A&A...430.1109A}
{Andries}, J., {Goossens}, M., {Hollweg}, J.~V., {Arregui}, I., \& {Van
  Doorsselaere}, T. 2005{\natexlab{b}}, \aap, 430, 1109

\bibitem[{{Andries} {et~al.}(2009{\natexlab{b}}){Andries}, {van Doorsselaere},
  {Roberts}, {Verth}, {Verwichte}, \& {Erd{\'e}lyi}}]{2009SSRv..149....3A}
{Andries}, J., {van Doorsselaere}, T., {Roberts}, B., {et~al.}
  2009{\natexlab{b}}, \ssr, 149, 3

\bibitem[{{Arregui} {et~al.}(2012){Arregui}, {Oliver}, \&
  {Ballester}}]{2012LRSP92A}
{Arregui}, I., {Oliver}, R., \& {Ballester}, J.~L. 2012, Living Reviews in
  Solar Physics, 9, 2

\bibitem[{{Asai} {et~al.}(2012){Asai}, {Ishii}, {Isobe}, {Kitai}, {Ichimoto},
  {UeNo}, {Nagata}, {Morita}, {Nishida}, {Shiota}, {Oi}, {Akioka}, \&
  {Shibata}}]{2012ApJ745L18A}
{Asai}, A., {Ishii}, T.~T., {Isobe}, H., {et~al.} 2012, \apjl, 745, L18

\bibitem[{Banerjee {et~al.}(2007)Banerjee, Erd\'{e}lyi, Oliver, \&
  O’Shea}]{banerjee}
Banerjee, D., Erd\'{e}lyi, R., Oliver, R., \& O’Shea, E. 2007, \solphys, 246,
  3

\bibitem[{Bogdan \& Judge(2006)}]{OASO}
Bogdan, T.~J. \& Judge, P. 2006, Phil. Trans. R. Soc. London, Ser. A, 364, 313

\bibitem[{Bonet {et~al.}(2005)Bonet, M{\'a}rquez, Muller, Sobotka, \&
  Roudier}]{bonet}
Bonet, J., M{\'a}rquez, I., Muller, R., Sobotka, M., \& Roudier, T. 2005, \aap,
  430, 1089

\bibitem[{Christopoulou {et~al.}(2000)Christopoulou, Georgakilas, \&
  Koutchmy}]{ORWS}
Christopoulou, E.~B., Georgakilas, A.~A., \& Koutchmy, S. 2000, \aap., 354, 305

\bibitem[{{Cooper} {et~al.}(2003{\natexlab{a}}){Cooper}, {Nakariakov}, \&
  {Tsiklauri}}]{2003A&A...397..765C}
{Cooper}, F.~C., {Nakariakov}, V.~M., \& {Tsiklauri}, D. 2003{\natexlab{a}},
  \aap, 397, 765

\bibitem[{{Cooper} {et~al.}(2003{\natexlab{b}}){Cooper}, {Nakariakov}, \&
  {Williams}}]{2003A&A...409..325C}
{Cooper}, F.~C., {Nakariakov}, V.~M., \& {Williams}, D.~R. 2003{\natexlab{b}},
  \aap, 409, 325

\bibitem[{{De Moortel}(2009)}]{2009SSRv..149...65D}
{De Moortel}, I. 2009, \ssr, 149, 65

\bibitem[{Dorotovi\v{c} {et~al.}(2008)Dorotovi\v{c}, Erd\'{e}lyi, \&
  Karlovsk\'{y}}]{doretala}
Dorotovi\v{c}, I., Erd\'{e}lyi, R., \& Karlovsk\'{y}, V. 2008, in Proc. IAU
  Symposium No. 247, ed. R.~Erd\'{e}lyi \& C.~A. Mendoza-Brice$\tilde{n}$o,
  Vol. 247 (Cambridge University Press), 351

\bibitem[{{Erd{\'e}lyi}(2008)}]{rbook}
{Erd{\'e}lyi}, R. 2008, in Physics Of The Sun And Its Atmosphere, ed. B.~N.
  Dwivedi \& U.~Narain (World Scientific Publishing)

\bibitem[{{Erd{\'e}lyi} {et~al.}(2013){Erd{\'e}lyi}, {Hague}, \&
  {Nelson}}]{2013SoPh..tmp..195E}
{Erd{\'e}lyi}, R., {Hague}, A., \& {Nelson}, C.~J. 2013, \solphys

\bibitem[{Fedun {et~al.}(2011{\natexlab{a}})Fedun, Shelyag, \&
  Erd{\'e}lyi}]{fedun2}
Fedun, V., Shelyag, S., \& Erd{\'e}lyi, R. 2011{\natexlab{a}}, \apj, 727, 17

\bibitem[{Fedun {et~al.}(2011{\natexlab{b}})Fedun, Shelyag, Verth,
  Mathioudakis, \& Erd{\'e}lyi}]{fedun1}
Fedun, V., Shelyag, S., Verth, G., Mathioudakis, M., \& Erd{\'e}lyi, R.
  2011{\natexlab{b}}, Ann. Geophys., 29, 1029

\bibitem[{Fujimura \& Tsuneta(2009)}]{fujimura}
Fujimura, D. \& Tsuneta, S. 2009, \apj, 702, 1443

\bibitem[{Gary(2001)}]{gary}
Gary, G. 2001, \solphys, 203, 71

\bibitem[{Goedbloed \& Poedts(2004)}]{goedbloed}
Goedbloed, J.~P. \& Poedts, S. 2004, Principles of magnetohydrodynamics: With
  applications to laboratory and astrophysical plasmas (Cambridge Univ Press)

\bibitem[{{Goossens}(2003)}]{2003ASSL..294.....G}
{Goossens}, M., ed. 2003, Astrophysics and Space Science Library, Vol. 294, {An
  introduction to plasma astrophysics and magnetohydrodynamics}

\bibitem[{{Goossens} \& {Poedts}(1992)}]{1992ApJ...384..348G}
{Goossens}, M. \& {Poedts}, S. 1992, \apj, 384, 348

\bibitem[{Huang {et~al.}(1998)Huang, Shen, Long, Wu, Shih, Zheng, Yen, Tung, \&
  Liu}]{huang}
Huang, N., Shen, Z., Long, S., {et~al.} 1998, Proc. R. Soc. A, 454, 903

\bibitem[{Jess {et~al.}(2009)Jess, Mathioudakis, Erd\'{e}lyi, Crockett, Keenan,
  \& Christian}]{jess}
Jess, D., Mathioudakis, M., Erd\'{e}lyi, R., {et~al.} 2009, Science, 323, 1582

\bibitem[{Khomenko {et~al.}(2008)Khomenko, Collados, \& Felipe}]{khomenko}
Khomenko, E., Collados, M., \& Felipe, T. 2008, \solphys, 251, 589

\bibitem[{Luna-Cardozo {et~al.}(2012)Luna-Cardozo, Verth, \&
  Erd\'{e}lyi}]{luna-cardozo}
Luna-Cardozo, C., Verth, G., \& Erd\'{e}lyi, R. 2012, \apj, 748, 110

\bibitem[{Malins \& Erd\'{e}lyi(2007)}]{malins}
Malins, C. \& Erd\'{e}lyi, R. 2007, \solphys, 246, 41

\bibitem[{Marsh \& Walsh(2008)}]{marsh2008p}
Marsh, M. \& Walsh, R. 2008, The Astrophysical Journal, 643, 540

\bibitem[{Mathew(2008)}]{mathew}
Mathew, S.~K. 2008, \solphys, 251, 515

\bibitem[{McAteer {et~al.}(2003)McAteer, Gallagher, Williams, Mathioudakis,
  Bloomfield, Phillips, \& Keenan}]{mcateer2003observational}
McAteer, R., Gallagher, P., Williams, D., {et~al.} 2003, \apj, 587, 806

\bibitem[{Moreels \& Van~Doorsselaere(2013)}]{moreels2013phase}
Moreels, M. \& Van~Doorsselaere, T. 2013, \AA, 551

\bibitem[{{Moreels} {et~al.}(2013){Moreels}, {Goossens}, \& {Van
  Doorsselaere}}]{michal2013}
{Moreels}, M.~G., {Goossens}, M., \& {Van Doorsselaere}, T. 2013, \aap, 555,
  A75

\bibitem[{{Morton} \& {Erd{\'e}lyi}(2009)}]{2009A&A...502..315M}
{Morton}, R.~J. \& {Erd{\'e}lyi}, R. 2009, \aap, 502, 315

\bibitem[{Morton {et~al.}(2011)Morton, Erd\'{e}lyi, Jess, \&
  Mathioudakis}]{morton}
Morton, R.~J., Erd\'{e}lyi, R., Jess, D.~B., \& Mathioudakis, M. 2011, \apj,
  729, L18

\bibitem[{{Morton} \& {Ruderman}(2011)}]{2011A&A...527A..53M}
{Morton}, R.~J. \& {Ruderman}, M.~S. 2011, \aap, 527, A53

\bibitem[{{Morton} {et~al.}(2012){Morton}, {Verth}, {Jess}, {Kuridze},
  {Ruderman}, {Mathioudakis}, \& {Erd{\'e}lyi}}]{2012NatCo...3E1315M}
{Morton}, R.~J., {Verth}, G., {Jess}, D.~B., {et~al.} 2012, Nature
  Communications, 3

\bibitem[{Pint\'er \& Erd\'elyi(2011)}]{pinter2011effects}
Pint\'er, B. \& Erd\'elyi, R. 2011, \ssr, 158, 471

\bibitem[{{Ruderman}(2003)}]{2003A&A...409..287R}
{Ruderman}, M.~S. 2003, \aap, 409, 287

\bibitem[{Ruderman \& Erd{\'e}lyi(2009)}]{ruderman2009transverse}
Ruderman, M.~S. \& Erd{\'e}lyi, R. 2009, Space Sci. Rev., 149, 199

\bibitem[{Rutten {et~al.}(2004)Rutten, Hammerschlag, Bettonvil, S{\"u}tterlin,
  \& De~Wijn}]{rutten}
Rutten, R., Hammerschlag, R., Bettonvil, F., S{\"u}tterlin, P., \& De~Wijn, A.
  2004, \aap, 413, 1183

\bibitem[{Scharmer {et~al.}(1985)Scharmer, Brown, Pettersson, \&
  Rehn}]{scharmer}
Scharmer, G., Brown, D., Pettersson, L., \& Rehn, J. 1985, Applied Optics, 24,
  2558

\bibitem[{Terradas {et~al.}(2004)Terradas, Oliver, \& Ballester}]{terradas}
Terradas, J., Oliver, R., \& Ballester, J. 2004, \apj, 614, 435

\bibitem[{Thompson(2006)}]{thompson2006magnetohelioseismology}
Thompson, M.~J. 2006, Phil. Trans. R. Soc. London, Ser. A, 364, 297

\bibitem[{Torrence \& Compo(1998)}]{torrence}
Torrence, C. \& Compo, G. 1998, Bulletin of the American Meteorological
  Society, 79, 61

\bibitem[{{Verth} \& {Erd{\'e}lyi}(2008)}]{2008A&A...486.1015V}
{Verth}, G. \& {Erd{\'e}lyi}, R. 2008, \aap, 486, 1015

\bibitem[{Vigeesh {et~al.}(2012)Vigeesh, Fedun, Hasan, \&
  Erd{\'e}lyi}]{vigeesh2012three}
Vigeesh, G., Fedun, V., Hasan, S., \& Erd{\'e}lyi, R. 2012, The Astrophysical
  Journal, 755, 18

\bibitem[{Wang(2004)}]{wang2004coronal}
Wang, T. 2004, in SOHO 13 Waves, Oscillations and Small-Scale Transients Events
  in the Solar Atmosphere: Joint View from SOHO and TRACE, Vol. 547, 417

\bibitem[{Wang(2011)}]{wang2011standing}
Wang, T. 2011, \ssr, 158, 397

\end{thebibliography}

\end{document}